\documentclass[fleqn,10pt]{article}
\usepackage[utf8]{inputenc}
\usepackage[T1]{fontenc}
\usepackage{graphicx}
\usepackage{siunitx}
\usepackage{authblk}
\usepackage{color,soulutf8}
\usepackage{amsmath}
\usepackage{hyperref}
\usepackage{verbatim}
\usepackage[square,numbers,sort&compress]{natbib}
\usepackage{ulem}
\title{Machine learning of microscopic structure-dynamics relationships in complex molecular systems}

\author[1]{Martina Crippa}
\author[2]{Annalisa Cardellini}
\author[1]{Matteo Cioni}
\author[3]{Gábor Csányi}
\author[1,2,*]{Giovanni M. Pavan}
\affil[1]{Department of Applied Science and Technology, Politecnico di Torino, Corso Duca degli Abruzzi 24, 10129 Torino, Italy}
\affil[2]{Department of Innovative Technologies, University of Applied Sciences and Arts of Southern Switzerland, Polo Universitario Lugano, Campus Est, Via la Santa 1, 6962 Lugano-Viganello, Switzerland}
\affil[3]{Engineering Laboratory, University of Cambridge, Trumpington Street, Cambridge, CB2 1PZ, United Kingdom}
\affil[*]{corresponding author: Giovanni M. Pavan (giovanni.pavan@polito.it)}
\date{}

\begin{document}


\maketitle
\begin{abstract}
In many complex molecular systems, the macroscopic ensemble's properties are controlled by microscopic dynamic events (or fluctuations) that are often difficult to detect via pattern-recognition approaches. 
Discovering the relationships between local structural environments and the dynamical events originating from them would allow unveiling microscopic-level structure-dynamics relationships fundamental to understand the macroscopic behavior of complex systems. 
Here we show that, by coupling advanced structural (\textit{e.g.}, Smooth Overlap of Atomic Positions, SOAP) with local dynamical descriptors (\textit{e.g.}, Local Environment and Neighbor Shuffling, LENS) in a unique dataset, it is possible to improve both individual SOAP- and LENS-based analyses, obtaining a more complete characterization of the system under study.
As representative examples, we use various molecular systems with diverse internal structural dynamics. On the one hand, we demonstrate how the combination of structural and dynamical descriptors facilitates decoupling relevant dynamical fluctuations from noise, overcoming the intrinsic  limits of the individual analyses.
Furthermore, machine learning approaches also allow extracting from such combined structural/dynamical dataset useful microscopic-level relationships, relating key local dynamical events (\textit{e.g.}, LENS fluctuations) occurring in the systems to the local structural (SOAP) environments they originate from.
Given its abstract nature, we believe that such an approach will be useful in revealing hidden microscopic structure-dynamics relationships fundamental to rationalize the behavior of a variety of complex systems, not necessarily limited to the atomistic and molecular scales.
\end{abstract}

\section*{Introduction}
The macroscopic behavior of complex systems is often influenced by fluctuations that, while being fundamental for comprehending the systems' dynamics, are challenging to detect and control.
This also holds true at the molecular scale, where phenomena such as nucleation, defect propagation, and phase transitions are intricately linked to these fluctuations.
The integration of advanced molecular descriptors with Machine Learning (ML) has been playing a key role in analyzing molecular trajectories, contributing to a better understanding of diverse nanoscale systems, ranging from atomistic to supramolecular levels.\cite{andrews2022,gasparotto2014, davies2022, noe2019, gardin2022, cardellini2022,capelli2021, lionello2022, cioni2022, rapetti2022, cheng2020}
Standard human-based descriptors, tailored for building detailed analyses and investigating specific systems like, i.e., ice-water interfaces \cite{errington2001} or metal clusters \cite{behler2007,rossi2018}, have increasingly left more and more space to abstract descriptors, \cite{bartok2013, pietrucci2015, behler2011,drautz2019,faber2015,gasparotto2020,musil2021} often combined with supervised and unsupervised ML methods.\cite{andrews2022,gasparotto2014, davies2022, noe2019, gardin2022, cardellini2022,capelli2021, lionello2022, cioni2022, rapetti2022}
These ML-based techniques offer valuable insights into the structural and dynamical properties of the systems. \cite{gardin2022, cardellini2022, capelli2021, lionello2022, cioni2022, rapetti2022}
While human-based approaches provide an accurate comprehension of intrigued physical-chemical mechanisms, they heavily rely on in-depth prior knowledge of the system, limiting their transferability.
On the contrary, the use of abstract descriptors allows more general representations and outlines a broader picture of the system behavior, eventually  managing a large amount of high-dimensional data which are often difficult to rationalize.
Widely recognized approaches based on dimensionality reduction principles (\textit{e.g.} linear Principal Component Analysis (PCA), kernel-PCA \cite{scholkopf1998}, t-SNE\cite{vandermaaten2008}), are frequently employed to extract information from such descriptors related dataset, then classifying the reduced dataset with diverse clustering methods (\textit{e.g.} KMeans\cite{lloyd1982}, Gaussian Mixture Models (GMM) \cite{reynolds2009}, DBSCAN\cite{schubert2017}, HDBSCAN \cite{McInnes2017}) to facilitate its interpretation.
However, when relying on structure-based descriptors, these approaches have limitations: while they effectively detect dominant structural environments in the system, they may fail to capture local time-dependent events that are sparsely observed within the trajectory.
These transitions, although statistically insignificant, have revealed a crucial role in the overall behavior of the system.\cite{crippa2022de,caruso2023}
The absence of an adaptive resolution that allows to catch non-dominant events presents two challenges: firstly, it leads to a loss of information by failing to detect fluctuations within the system, and secondly, these fluctuations may be inaccurately classified within the dominant clusters, thereby contaminating them.

In recent studies, we have developed dynamic descriptors that are very efficient in capturing the local dynamic environments of atoms in complex molecular systems from structural information/identity-based information. \cite{crippa2022de,caruso2023}
By monitoring these descriptors over time along the trajectory, we can effectively capture dynamic behaviors, including local and sparse events within the system.
In particular, we introduced a dynamical descriptor, LENS (Local Environments and Neighbors Shuffling)\cite{crippa2022de}, which considers the interacting particles as distinct individuals (IDs) monitoring how much the list of neighbor particles ID (of each particle \textit{i}) changes over time, for example at each sampled $\Delta$t.
As follows, LENS provides information on the reshuffling of the local neighbor's environments surrounding each unit \textit{i} in the system along the trajectory.
However, at the same time, a descriptor like LENS retains very limited structural information: if, \textit{e.g}, the neighbor units rearranging locally, while remaining within the neighborhood in $\Delta$t, LENS would not detect any signal (such events are vice-versa well captured by structural descriptors such as tSOAP\cite{caruso2023}).
Thus, while LENS can detect the local dynamics of the system, it does not allow to determine, \textit{e.g}, the specific structural environment from which dynamic events originate.

Here we demonstrate how, combining structural (SOAP \cite{bartok2013}) and dynamical (LENS \cite{crippa2022de}) descriptors, it is possible to obtain an improved characterization of the system.
We compose a dataset where the SOAP spectra ($n$ components each) are augmented with the LENS descriptor (an additional dimension), leading to significant technical and scientific advantages.
Firstly, (i) it enables the separation of sparsely observed, but relevant, dynamic events/environments (\textit{e.g.} fluctuations) from the noise in the SOAP dataset.
As a result, (ii) the interpretation of SOAP and LENS (combined) not only provides a more accurate complete characterization, but the two descriptors improve each other: the addition of LENS yields an enhanced SOAP structural classification.
Furthermore, (iii) this allows identifying unique microscopic structure-dynamics relationships, showing \textit{e.g.} which local SOAP structural environments generate a certain type of dynamical event along the sampled Molecular Dynamics (MD) trajectory.
In this work we demonstrate the efficiency and abstraction of this approach on diverse molecular systems, employed herein as case studies.

\section*{Results}
\begin{figure}
    \centering
    \includegraphics[width=\columnwidth]{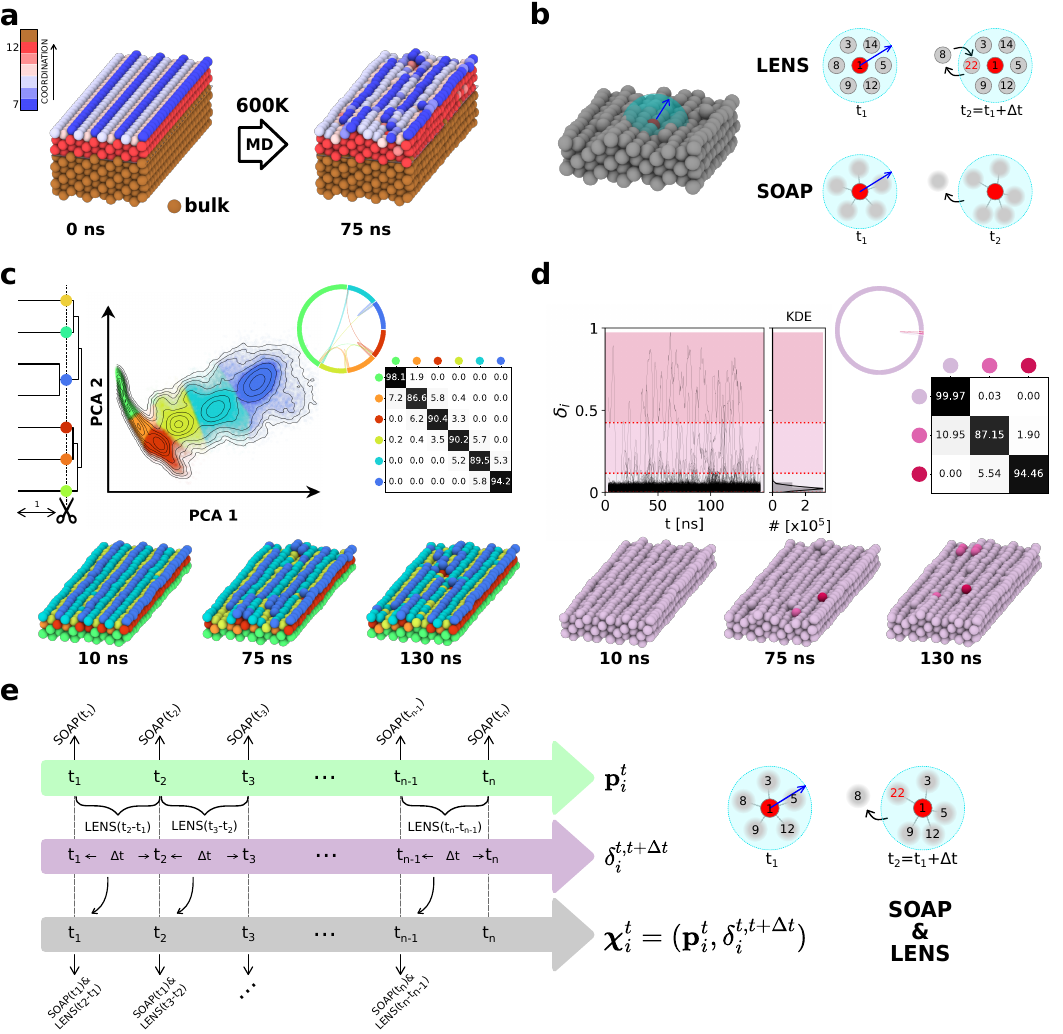}
    \caption{Flow of the analysis. (a) FCC 211 copper slab snapshots colored by atom coordination (excluding bulk) at t=0 \si{\nano\second} and after 75 \si{\nano\second} of an MD trajectory at T=600 \si{\kelvin}. (b) LENS and SOAP: given the local neighborhood (cyan sphere) of each atom (red atom) in the system, LENS tracks the \textit{identity} of the neighbor atoms within it (no information on their geometrical organization is retained), while SOAP captures their structural arrangement (without tracking their identity: it is a permutationally-invariant description). (c) SOAP-based analysis of \textbf{Cu(211)} system. Left: HC-based dendrogram (from an HDBSCAN* classification, see Figure S1a) and dendrogram cutting, defining the merged macro-clusters. Middle: PCA of the SOAP dataset (first two principal components), colored based on the detected macro-clusters. Right: chord diagram (fluxes) and transition probability matrix for the dynamical transitions between the macro-clusters (SOAP environments). Bottom: surface MD snapshots where atoms are colored based on the classification: bulk atoms in green, sub-surface in orange and red, surface "valleys" in yellow, faces in cyan, and edges in blue. (d) LENS analysis of \textbf{Cu(211)} at 600 K. Left: LENS time-series and classification.\cite{crippa2022de} Right: chord diagram (fluxes) and transition probability matrix. Bottom: MD snapshots with atoms colored based on the LENS clusters: more/less dynamic atoms in brighter/lighter colors. (e) Scheme of the SOAP\&LENS combined data set: the SOAP power spectrum of each particle at every time step ($\textbf{p}^t_i$) is combined with the LENS scalar value calculated at the subsequent time-interval ($\delta^{t+\Delta t}_i$), obtaining a new dataset $\boldsymbol{\chi}_i^{t}$.}
    \label{fig:01}
\end{figure}

As a first case study, we focus on a copper \textbf{Cu(211)} FCC metal surface recently demonstrated to possess non-trivial internal atomic dynamics.
Metals are known to display interesting dynamic behavior even well below the melting temperature.\cite{spencer1986, jayanthi1985}
For example, when simulated at T=600 $\si{\kelvin}$, the \textbf{Cu(211)} FCC slab of Figure \ref{fig:01} exhibits a surface with structurally diverse environments, as made evident by a simple coordination analysis, and a non-trivial internal atomic dynamics (Figure \ref{fig:01}a, right: dynamical atomic rearrangements).
Unveiling the underlying mechanism behind such dynamics is essential to understand the properties of such metal systems \cite{yamakov2004deformation, zepeda2017probing, wang2021atomistic}.
Moreover, the comprehension of structural-dependent features plays a fundamental role in practical applications such as heterogeneous catalysis, mechanical properties, \textit{etc}.\cite{koch1992, wang1991, antczak2010, gazzarrini2021}
SOAP-based and LENS-based ML analyses have been recently employed to analyze MD simulation trajectories of metals below the melting temperature (including, \textit{e.g.} copper surfaces as that of Figure \ref{fig:01}). \cite{cioni2022, crippa2022de, caruso2023}
Although a structural-descriptor-based analysis, such as that one using SOAP combined with dimensionality reduction and density-based clustering, captures the most prevalent and dominant conformation domains within the system, a pure LENS analysis based on the reshuffling of the neighborhood over time, catches the dynamical features of the system (see Figure \ref{fig:01}b).

Here, we investigate a \textbf{Cu(211)} FCC copper slab using a preexisting MD trajectory composed of N=2400 atoms simulated \textit{via} a DeepMD-based potential \cite{wang2018} for 150 \si{\nano\second} (see Cioni \textit{et al.} \cite{cioni2022} for details).
To examine both the structural and dynamical properties of the \textbf{Cu(211)} system, we firstly adopted a similar \textit{bottom-up} protocol as described in the study by Cioni \textit{et al.} \cite{cioni2022}.
This strategy includes, as a first step, a representation of the system via the SOAP descriptor.
One SOAP spectrum is extracted for each of the 900 atoms (three top-most layers, although the SOAP spectra also consider the presence of the 1500 bottom-side atoms as neighbors, they are not included in the analysis because we are interested in the dynamics of the surface \cite{cioni2022}) in 482 snapshots taken every $\Delta$t=0.3 \si{\nano\second} along 145 \si{\nano\second} of MD simulation, for a total of 482$\times$900 spectra.
A Principal Component Analysis (PCA) is then used to reduce the dimensionality of the SOAP spectra dataset, considering the first n-PCA components in order to retain at least 99.5\% of the variance (see Table S1 in Supporting Information for details).
Unsupervised clustering algorithm (HDBSCAN*\cite{McInnes2017} or Gaussian Mixture Models \cite{reynolds2009}) can be finally adopted to rationalize the data and to identify the dominant Atomic Environments (AE) on the surface (colored clusters in the PCA of Figure S1).
From the atoms' transition between the clusters over time, we compute a transition probability matrix. This reports the probability of an atom that is in a certain cluster at time $t$ to remain in the same environment at time t+$\Delta$t (\textit{i.e.}, after $\Delta$t: the temporal resolution of our analysis) or to undergo transition into a different cluster (see Figure S1 for the micro-clusters transition matrix).
From the transition probability matrix, we construct a Hierarchical Clustering (HC) based dendrogram merging the clusters with high dynamic interconnection (Figure S1).
The dendrogram is cut in order to retrieve only meaningful clusters, colored accordingly in the PCA plot of Figure \ref{fig:01}c, where only the first two PCA components are reported.
The results demonstrate how SOAP can successfully distinguish diverse structural environments within this system, including the bulk (green), subsurface (orange and red), surface valleys (yellow), faces and edges (cyan and blue), identified in different colors in Figure \ref{fig:01}c.
The dynamic interconnections between the various clusters (AEs) on the surface are also represented by the cord diagrams in Figure \ref{fig:01}c on the right: in these cord diagrams the dimensions of the arcs stand for the population of the various clusters, while the dimensions of the strings connecting them give visual information on how pronounced the atomic flow is in $\Delta$t, and thus on their dynamic interconnection.
Moreover, we also obtained the transition probabilities matrix (\% to undergo transition in $\Delta$t=0.3 \si{\nano\second}) between the HC-merged clusters (Figure \ref{fig:01}c right).

Separately, we also perform a LENS analysis on the same 482 snapshots extracted by the same MD trajectory.
A LENS analysis of the system reveals intriguing surface events that are not captured (or highlighted) by the static SOAP-based analysis of structure as described above.
Specifically, a few Cu atoms are seen to detach from the crystalline structure of the \textbf{Cu(211)} surface and to diffuse on it very fast.
On the one hand, since these diffusing atoms are characterized by a high rate of reshuffling of their neighbors, they are clearly identified by LENS as a separate environment in the dataset (Figure \ref{fig:01}d).
On the other hand, a comparison of Figure \ref{fig:01}c and Figure \ref{fig:01}d shows how, since these points are sparse and have negligible statistical weight in the dataset, these are overlooked in a pattern recognition approach such as that, \textit{e.g.}, of Figure \ref{fig:01}c.
In particular, in the SOAP analysis of Figure \ref{fig:01}c, it is possible to note that the diffusing atoms (magenta in Figure \ref{fig:01}d), are merged to the SOAP cluster identifying the edges of \textbf{Cu(211)} surface.
To address this limitation, here we developed a combined approach based on the basic assumption that a structural environment at a certain time might influence the  dynamical events within the subsequent time interval.
As shown schematically in Figure \ref{fig:01}e, starting for example at time $t_1$, a SOAP spectrum $\textbf{p}_i^{t_1}$ is computed for each particle $i$ in the system.
We also calculate its LENS value for the immediately subsequent time interval $\delta_i^{t_2 - t_1}$.
By including the LENS term as an extra-component into each SOAP power spectrum, we thus obtain a new vector $\boldsymbol{\chi}_i^{t_1}=(\textbf{p}_i^{t_1}, \delta_i^{t_2 - t_1})$ containing information on the structural properties in the neighbor environment surrounding atom $i$ at time $t_1$ and its evolution in the subsequent time interval $t_2-t_1$.
The SOAP spectrum and LENS scalar component are opportunely normalized to have the same weight in the dataset (see Methods for details).
Iterating this procedure for the whole trajectory, we thus obtain a new dataset (SOAP\&LENS dataset) comprising $N = N_{particle}\times N_{frames}$ vectors, each one of dimension $n+1$, where $n$ is the SOAP spectrum dimension (structural information) and 1 the LENS (dynamical) component.
Such updated dataset effectively contains information on the instantaneous environments surrounding each particle $i$ and how they are prone to change over time at the resolution $\Delta$t (0.3 \si{\nano\second}) of our analysis.

\begin{figure}
    \centering
    \includegraphics[width=\columnwidth]{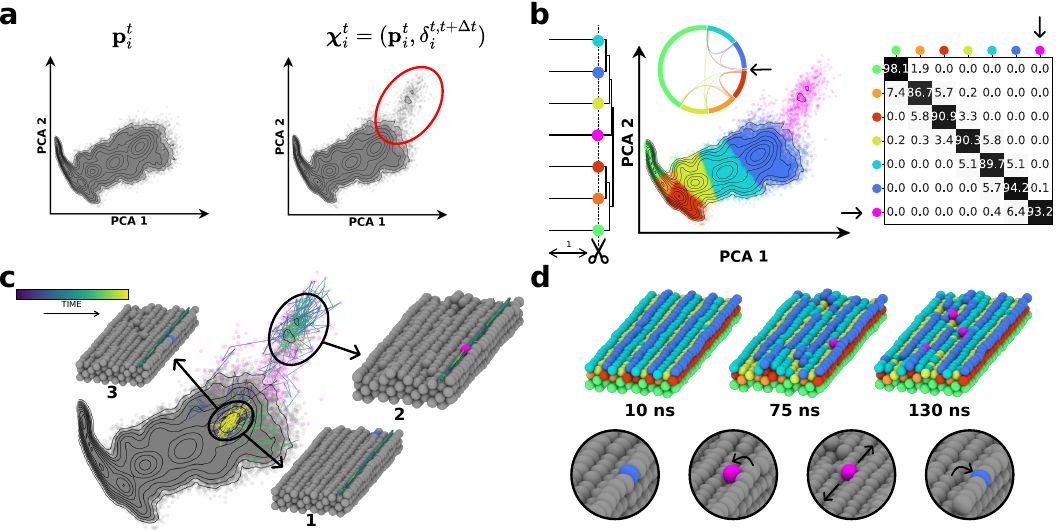}
    \caption{Combined SOAP\&LENS analysis of a Cu(211) surface at 600 K. (a) Left: First two PCA components of the SOAP power spectra of the \textbf{Cu(211)} system. Right: First two PCA components of the SOAP\&LENS combined descriptor: the new cloud of points emerging in the PCA projection of the $\boldsymbol{\chi}$ vector is highlighted by the red circle. (b) SOAP\&LENS based analysis of \textbf{Cu(211)} system. Left: HC-based dendrogram (from an HDBSCAN* classification, see Figure S2a) and dendrogram cutting, defining the merged macro-clusters (accordingly to clusters in Figure \ref{fig:01}a, except for a new pink cluster). Middle: PCA of the SOAP\&LENS dataset (first two principal components), colored based on the detected macro-clusters and chord diagram (fluxes). Right: transition probability matrix for the dynamical transition between the macro-clusters, highlighting the new cluster in pink. (c) Trajectory of an atom detaching from an edge, running on the surface, and re-entering into the edge. The trajectory is shown both on the PCA plot and on the snapshots, colored from blue to yellow according to the time evolution.
    (d) Three surface MD snapshots colored based on the classification: bulk atoms in green, sub-surface in orange and red, surface "valleys" in yellow, faces in cyan, edges in blue and pink atoms detaching from the edges and running on the surface (an example of this process is reported in the zoom below).}
    \label{fig:02}
\end{figure}

This method allows us to delineate a new concept for classification, as reported in Figure \ref{fig:02}.
On the left side, Figure \ref{fig:02}a shows the PCA of the SOAP dataset projected onto the first two PCA components.
On the right side, Figure \ref{fig:02}a shows the same projection for the new SOAP\&LENS combined dataset (see Methods section for additional details).
Notably, while the majority of the data has an almost identical distribution on the two PCAs, a distinct cloud of points appears as evidently separated from the rest in the combined dataset (top-right: highlighted by the red circle).
Shown in Figure \ref{fig:02}b, unsupervised HDBSCAN* clustering combined with HC-based merging (in general, any other suitable clustering algorithm, \textit{e.g.} GMM, DBSCAN, KMeans, would also work) reveals that such a separated domain on the SOAP\&LENS PCA identifies a distinct, specific local environment.
Note that the clustering parameters used for the analyses of Figures \ref{fig:01}c and \ref{fig:02}b are exactly the same (see Methods for details).
This comparison shows how the classification of Figure \ref{fig:01}c (SOAP only) is enriched via the detection of a new LENS environment identified by the pink color (highlighted by the arrows in the transition matrix and chord diagram of Figure \ref{fig:02}b).
As done for both the SOAP and LENS independent analyses, we can reconstruct the evolution of the detected environments by following the AE belonging to all atoms at every time step (see the chord diagram and transition probability matrix in Figure \ref{fig:02}b, right).

This analysis based on combining SOAP and LENS in a unique dataset offers distinct advantages over the purely SOAP-based approach.
The decoupling of this additional pink LENS environment not only provides a more complete description of what happens in the \textbf{Cu(211)} surface at 600 \si{\kelvin}, but also improves the statistical precision in the classification of the SOAP environments.
In fact, in differentiating the structural from the dynamical environments, the detection of the SOAP AEs in the SOAP\&LENS dataset benefits from a reduced error.
Notably, the PCA area identified by the red oval in Figure \ref{fig:02}b, which corresponds in this analysis to a well-defined LENS AE, merges into the SOAP AEs in the PCA of Figure \ref{fig:01}c, creating errors and increased uncertainty.
In this sense, when combined, two distinct descriptors such as, \textit{e.g.}, SOAP and LENS, complement and improve each other.
Furthermore, such an approach also allows tracking the origin of local dynamical (LENS) fluctuations occurring on the surface, outlining microscopic structure-dynamics relationships.
The off-diagonal entry in the matrix of Figure \ref{fig:02}b representing the transition of atoms from the edge AE (in blue) to the pink (LENS) environment ($\sim$0.1 \% probability) reveals that those atoms diffusing with high-speed on the metal surface come from the surface edges (see Movie S1).
After their creation and diffusion, such diffusing pink atoms are then again reabsorbed into the surface edges ($\sim$6.4 \% probability).
The large imbalance between the probabilities for the creation and annihilation of these LENS diffusing atoms (Figure \ref{fig:02}b right, $\sim$0.1\% vs. $\sim$6.4\%) indicates that the emergence of such fast atoms is a rare event.
Yet, it is clear that detecting such diffusing atoms is key for understanding the behavior of the system.
Figure \ref{fig:02}c provides an example of the structural variation of an atom undergoing such transition, following its trajectory both on the PCA plot and along the MD.
The atom's trajectory is color-coded based on the MD simulation time, from dark blue to yellow, showing atoms that after residing within the surface edges (dark blue lines, example snapshot 1), detach and diffuse on the surface becoming part of this pink LENS environment (green lines, example snapshot 2), and then being reabsorbed into the edges (yellow lines, example snapshot 3).
Figure \ref{fig:02}d shows a complete representation of the \textbf{Cu(211)} surface colored based on corresponding SOAP\&LENS environments.
In contrast to the snapshots of Figure \ref{fig:01}c,d, this comprehensive approach captures all the key SOAP as well as LENS environments, providing a more complete characterization of this system.

By combining these two descriptors, it becomes evident that the motion of atoms diffusing on the surface (pink AE) originates from fluctuations within the SOAP environment, which defines the edges of the surface (blue AE).

\begin{figure}
    \centering
    \includegraphics[width=\columnwidth]{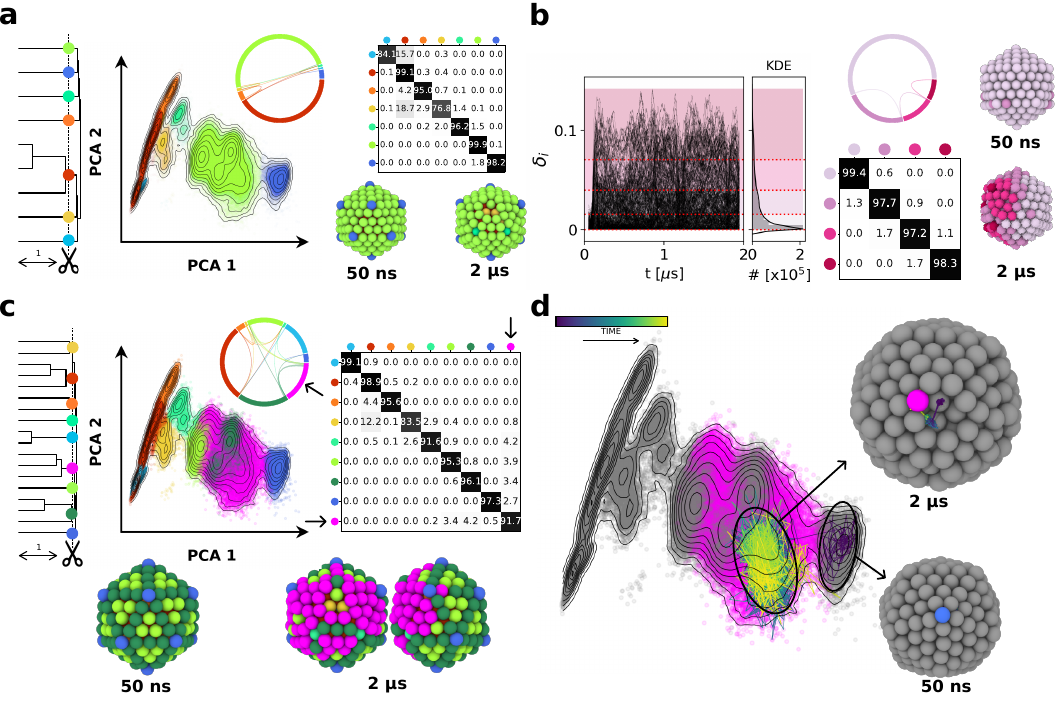}
    \caption{Combined SOAP\&LENS analysis of a cold Au-NP surface at 200 K. (a) SOAP-based analysis of the \textbf{Au-NP} system. Left: HC-based dendrogram (from an HDBSCAN* classification, see Figure S1b) and dendrogram cutting, defining the merged macro-clusters. Middle: PCA of the SOAP dataset (first two principal components), colored based on the detected macro-clusters, and chord diagram (fluxes). Right: transition probability matrix for the dynamical transitions between the macro-clusters (SOAP environments). Bottom: two nanoparticle MD snapshots where atoms are colored based on the classification: vertices in blue, surface atoms in lime, sub-surface in light-green, yellow and orange, and bulk atoms in red. (b) LENS analysis of \textbf{Au-NP}. Left: LENS time-series and classification. Middle: chord diagram (fluxes) and transition probability matrix. Right: MD snapshots with atoms colored based on the LENS clusters: more/less dynamic atoms in brighter/lighter colors. (c) SOAP\&LENS based analysis of the \textbf{Au-NP}. Left: HC-based dendrogram (from an HDBSCAN* classification, see Figure S2b) and dendrogram cutting, defining the merged macro-clusters, colored according to cluster classification in Figure \ref{fig:03}a, except for a new pink cluster. Middle: PCA of the SOAP dataset (first two principal components), colored based on the detected macro-clusters, and chord diagram (fluxes). Right: transition probability matrix for the dynamical transitions between the macro-clusters, highlighting the new cluster in pink. Bottom: three nanoparticle MD snapshots colored according to the classification: vertices in blue, surface atoms in lime, sub-surface in light-green, yellow and orange, bulk atoms in red  and "liquid-like" region in pink.  (d) Trajectory of an atom detaching from a vertex and entering the surface of the nanoparticle and giving rise to the rosette environment. The trajectory is shown both on the PCA plot and on the snapshots, colored from blue to yellow according to the time evolution.}
    \label{fig:03}
\end{figure}

We  further test our approach on different systems.
We carried out a second test on a 309 atoms icosahedral gold nanoparticle (\textbf{Au-NP}) model, simulated for 2 $\si{\nano\second}$ at T=200 $\si{\kelvin}$ using the Gupta potential \cite{gupta1981, rapetti2022}, (see Methods section for details). In these conditions, this \textbf{Au-NP} was demonstrated to have non-trivial dynamics \cite{rapetti2022,crippa2022de}.
In Figure \ref{fig:03}a, a SOAP-based analysis of the MD trajectory reveals the dominant structural environments within the NP vertices in blue, surface in lime, sub-surfaces AEs in orange, bulk atoms in red and also surface defects in yellow and rosette in light-green.
The dynamics of these SOAP AEs is quantified by the exchange chord diagram and in the transition probability of Figure \ref{fig:03}a (right).
At the same time, analysis of the LENS time series unveils a crucial insight, overlooked by a pure SOAP analysis (Figure \ref{fig:03}b).
After $\sim$ 180 \si{\nano\second} of MD simulation, the nanoparticle undergoes a sharp local structural transition involving one vertex, which penetrates the surface  generating a distinctive structure known as a rosette (Figure \ref{fig:03}a,d: in light-green).
Notably, the creation of a rosette (six symmetrical neighbors around an intruded center) from a vertex (five symmetrical neighbors) is an event that is known to happen in such icosahedral NPs and that can be observed experimentally.\cite{rapetti2022, apra2004}.
The LENS analysis shows the emergence of strong signals when the vertex intrudes and triggers the formation of the rosette (Figure \ref{fig:03}b, left).
In particular, the magenta colors in Figure \ref{fig:03}b reveal, after such local transition, the presence of a highly dynamic "liquid-like" region surrounding the rosette, coexisting with a "crystalline-like" domain in the remaining portion of the \textbf{Au-NP}.
It is worth noting how a SOAP analysis alone overlooks such a dynamic surface non-uniformity: for the SOAP descriptor, rich in structural information, this local dynamical change does constitute a relevant effect.
In the SOAP-based analysis, such a "liquid-like" region is classified together with the crystalline region, as a global surface cluster (lime color), even if the dynamic behavior of the two regions is different.
Therefore, the SOAP description fails to capture part of the system physics: it incorporates two distinct regions with entirely different dynamical behaviors into one single cluster characterized by an averaged structural representation.

In Figure \ref{fig:03}c, we show the results of SOAP\&LENS based analysis, where we combined the SOAP spectrum of each atom at every timestep with the LENS signal for the same atom at the subsequent $\Delta$t.
In this case, the combined analysis reveals that a significant portion of the PCA-reduced data -in particular, that central region referring to the surface of the NP (in Figure \ref{fig:03}a: in lime)- corresponds to a highly dynamic LENS environment (Figure \ref{fig:03}c: pink).
This allows us to disentangle the "liquid-like" region from the well-defined crystal-like structural domains on the NP surface.
Furthermore, the results of Figure \ref{fig:03}c demonstrate again how, also in this case, the addition of LENS improves the accuracy in the detection of the SOAP environments.
Comparing of Figure \ref{fig:03}a \textit{vs.} \ref{fig:03}c, it is clear how the analysis robustly distinguishes now the edges (dark green), faces (lime) and vertexes (blue), as well as rosettes (light-green) and defects (yellow) on the icosahedral NP surface.
Similar to the case of \textbf{Cu(211)}, a strong correlation arises between the "liquid-like" dynamical domain and specific structural environments: the LENS (pink) cluster in the transition matrix is found connected to the faces (lime, $\sim$ 3.9 \%), edges (green, $\sim$ 3.4 \%), vertices (blue, $\sim$ 2.7 \%) and especially the rosettes (light-green, $\sim$ 4.2) of the NP.
This is interesting, considering that the pink dynamical region (local "melting" of the NP surface) originates from the creation of a first rosette (a defect in the icosahedron).

In Figure \ref{fig:03}d, we show an example of a structural variation event that gives rise to the formation of a rosette structure.
This transition is depicted both on the PCA plot and on the snapshots, where the trajectory of the vertex atom (blue, at 50 \si{\nano\second}) is color-coded according to time evolution, ranging from dark blue to yellow (2 \si{\micro\second} of MD).
This demonstrates how the vertex atom entering into the surface, leads to the emergence of a "liquid-like" region surrounding the rosette (pink, at 2 \si{\micro\second} of MD).

\begin{figure}
    \centering
    \includegraphics[width=\columnwidth]{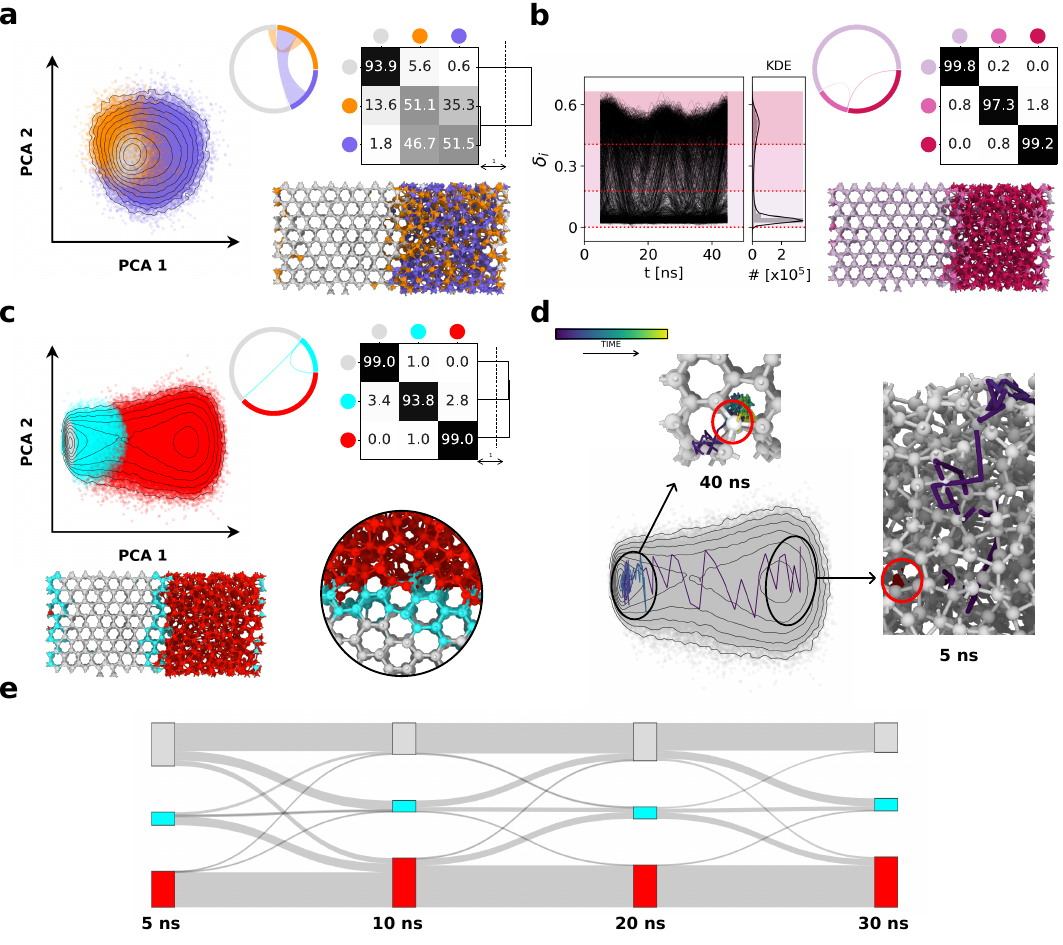}
    \caption{Combined SOAP\&LENS analysis of ice-liquid water equilibrium in correspondence of the transition temperature. (a) SOAP-based analysis of the \textbf{TIP4P/Ice} system. Left: first two PCA plot, colored according to GMM clustering (see Methods section for details). Right: chord diagram (fluxes) and transition probability matrix of the clusters and its HC-based dendrogram showing the relation within them. Bottom: a snapshot along the trajectory colored based on the cluster classification, ice molecules in white, liquid water in orange and purple. (b) LENS-based analysis of \textbf{TIP4P/Ice}. Left: LENS time-series and classification. Middle: chord diagram (fluxes) and transition probability matrix. Bottom: an MD snapshot with atoms colored based on the LENS clusters, more/less dynamic atoms in brighter/lighter colors. (c) SOAP\&LENS based analysis of the \textbf{TIP4P/Ice}. Left: first two PCA plot, colored according to GMM clustering (see Methods section for details). Right: the chord diagram (fluxes) and transition probability matrix of the clusters and its HC-based dendrogram showing the relation within them. Bottom: an MD snapshot colored according to the cluster classification: ice molecules in white, liquid water in red and the interface in cyan. Right: zoom of the interface region. (d) Trajectory of a molecule that undergoes a phase transition, from liquid water to ice, crossing the interface. The trajectory is reported both on the PCA plot and on the zoomed snapshots, colored from blue to yellow according to the time evolution. (e) Flow chart of the transitions between the three phases, colored accordingly, at 5 $\si{\nano\second}$, 10 $\si{\nano\second}$, 20 $\si{\nano\second}$ and 30 $\si{\nano\second}$.}
    \label{fig:04}
\end{figure}

As a last case study, we present the effectiveness of our SOAP\&LENS analysis in capturing distinct phases within a system where ice and liquid water coexist in correspondence of the solid and liquid transition temperature.
We analyzed 50 \si{\nano\second} of an atomistic simulation of water modeled with \textbf{TIP4P/Ice} force field, containing 2048 molecules at equilibrium between the two phases ($\sim$50\% ice and $\sim$50\% liquid water) at the transition temperature.\cite{crippa2022de, caruso2023}
A pure SOAP-based (structural) analysis, reported in Figure \ref{fig:04}a, can distinguish the two main phases (ice in white and liquid water in orange and purple).
The two clusters in orange and purple in Figure \ref{fig:04}a, correspond to tiny variations of the same environment (liquid water).
This is clearly shown in the probability matrix and in particular in the HC-based dendrogram, where the purple and orange AEs are very close to one each, and both are in comparison very far from the white one (see Figure \ref{fig:04}a right).
However, recently we have demonstrated that a pure LENS (dynamic) analysis can detect easily both the ice and water environments, plus also the interface between them.\cite{crippa2022de}
Figure \ref{fig:04}b shows the LENS time series, which clearly highlight two distinct statistically relevant environments, with different dynamics, separated by an interface environment where the ice/liquid water molecular transitions occur.
The flux chord diagram and the probability transition matrix of Figure \ref{fig:04}b (right) reveal how the ice/liquid phase transition of the molecules takes place through the interface.
Figure \ref{fig:04}c displays the combined SOAP and LENS in a unique dataset, thereby providing a PCA that is significantly distorted compared to the SOAP one of Figure \ref{fig:04}a.
Two main density peaks are evident (in white and red) corresponding respectively to ice and liquid water. 
GMM clustering now clearly detects a distinct area on the PCA corresponding to the ice-water interface (in cyan).
In Figure \ref{fig:04}d, we highlighted one explicatory trajectory (on the PCA plot and on the snapshot) of a water molecule undergoing phase transition from liquid water to ice, crossing the interface. 
The flow chart in Figure \ref{fig:04}e provides a qualitative visualization of the transitions between the various environments considering specific time intervals along the trajectory (\textit{e.g.} at 5 $\si{\nano\second}$, 10 $\si{\nano\second}$, 20 $\si{\nano\second}$ and 30 $\si{\nano\second}$).
Also in this case, the addition of a LENS component to the SOAP vectors offers a clear advantage over a purely structural analysis (SOAP only).
In this specific case, it is interesting to note how LENS retains large part of the information contained in the system trajectory compared to SOAP.
This is evident, for example, if we compare the cumulative variance contained in the dataset as a function of the number of principal components of the PCA.
In Figure S4, we clarify that to reach the 99\% of the cumulative variance of the dataset 8 components are needed in a purely structural SOAP dataset, while for example, when LENS is embedded into the dataset, with only three components we largely exceed the 99\% of variance.
This demonstrates how, in this system, the LENS descriptor might retain more comprehensive information regarding the key features that characterize the system, compared to the SOAP descriptor.
 
In conclusion, this study points out the intrinsic limitations of relying solely on structural descriptors to comprehend the physics of dynamically evolving systems.
By integrating a microscopic dynamic descriptor, like LENS, with a structural counterpart (\textit{e.g.} SOAP), we obtain numerous advantages.
First, this integration improves the accuracy of both structural and dynamic classifications, ``cleaning up'' the noise and reducing the degeneracy issues intrinsic to both individual analyses.
Second, this paves the way for understanding how given structural microscopic environments within the system can generate specific dynamic behavior (fluctuations).
This opens new routes to learn microscopic-scale structure-dynamic relationships (\textit{e.g.} those of Figures \ref{fig:02} and \ref{fig:03}) that are key to understanding the behaviors and properties of these, and in general of a variety of complex systems.
These results are also reminiscent of general concepts in physics.
For instance, when studying the behavior of a system, the sole positional information of the objects is insufficient to predict the dynamic behavior of the system at non-zero temperature (\textit{e.g.}, information on velocities is also needed).
Similarly, these results demonstrate how coupling a purely structural parameter like SOAP, which provides information only on the relative structural arrangements, with a descriptor that is rich in local dynamic information, offers fascinating insights.
We expect that such type of approach, given its abstract nature, will be highly valuable in characterizing the behavior of complex systems across various domains and potentially also beyond the atomistic/molecular scale.

\section*{Methods}
\subsection*{MD simulations and pre-processing}
The atomistic model of \textbf{Cu(211)} surface (see Figures \ref{fig:01},\ref{fig:02}) is composed of N$_{211}$=2400 atoms. The MD simulation is conducted at T=600 \si{\kelvin} \textit{via} LAMMPS software\cite{thompson2022} using a Neural Network potential built using the DeepMD platform,\cite{wang2018} as described in detail in reference \cite{cioni2022}.
The sampled trajectories are 150 \si{\nano\second} long. A total of 502 frames are extracted every $\Delta$t=0.3 \si{\nano\second} along the MD trajectory and used for the analysis.

The atomistic model for the icosahedral \textbf{Au-NP} is composed of N$_{Au-NP}$=309 gold atoms (Figure \ref{fig:03}). The \textbf{Au-NP} model is parametrized according to the Gupta potential, \cite{gupta1981} and is simulated for 2 \si{\micro\second} of MD at T=200 \si{\kelvin} using the LAMMPS software\cite{thompson2022} as described in detail in reference \cite{rapetti2022}.
2000 frames are extracted every $\Delta$t=1 \si{\nano\second} of the MD trajectory and then used for the analysis.

The atomistic Ice/Water interface model of Figure \ref{fig:04} is composed of N$_{TIP4P}$=2048 water molecules. The MD simulation is conducted at T=268 $\si{\kelvin}$. The \textbf{TIP4P}/\textbf{Ice} water model\cite{abscal2005} is used to represent both the solid phase of ice and the phase of liquid water,\cite{wang2018} as described in detail in reference \cite{crippa2022de}.
The sampled trajectory is t=50 \si{\nano\second} long, sampled and analyzed every 0.1 \si{\nano\second}.

All MD trajectories are firstly pre-processed in order to obtain a \texttt{hdf5} database, containing the data needed to extract the SOAP spectra and LENS values by using the software SOAPify, accessible at: \url{https://github.com/GMPavanLab/SOAPify}.
For the \textbf{Cu(211)} surface, we computed the SOAP spectra on both the surface and the bulk (N$_{211}$=2400 atoms in total), removing most of the deep bulk atoms, thus obtaining the 900-atoms system analyzed herein.
We analyzed all the atoms of the \textbf{Au-NP} system.
In the \textbf{TIP4P/Ice} water system of Figure \ref{fig:04} we computed the SOAP spectra for all the O atoms considering also the H atoms in the environment, while we did the LENS analysis by considering only the O atoms.
In all cases, the analysis is then conducted by building both the local SOAP environments and the LENS values of each unit within a sphere of radius r$_{cut}$ (see \ref{fig:01}b), equal to 6 $\si{\angstrom}$ for the \textbf{Cu(211)}, 4.48 $\si{\angstrom}$ for the \textbf{Au-NP}, and 6 $\si{\angstrom}$ for the \textbf{TIP4P/Ice} system.

\subsection*{SOAP analysis}
To describe the structural environment surrounding each particle within the simulations, we use the SOAP descriptor.
We compute the SOAP spectrum \textbf{p}$_i^t$ representing the local structural environment of each particle $i$ at every timestep $t$ within a cut-off radius r$_{cut}$ (6 $\si{\angstrom}$ for the \textbf{Cu(211)}, 4.48 $\si{\angstrom}$ for the \textbf{Au-NP}, and 6 $\si{\angstrom}$ for the \textbf{TIP4P/Ice} system) through the software SOAPify, accessible at: \url{https://github.com/GMPavanLab/SOAPify}. 
The SOAP vectors are generated using \textit{dscribe}\cite{HIMANEN2020}, and both l$_{max}$ and n$_{max}$ parameters for spherical harmonics, and number of radial basis functions are set to 8.
The results in a 576-component vector represent the environments of one particle at a certain timestep for the single species systems in (\textbf{Cu(211)} and \textbf{Au-NP}), while in a 1728-component vector for the ice/water interface.
Then, we applied the Principal Component Analysis algorithm to each dataset (as implemented in the SciPy python package  \cite{virtanen2020}), reducing the dimensionality of the representation to the first n-components, in order to reach a certain cumulative variance within each system, as reported in Table S1.
To analyze the reduced data of the \textbf{Cu(211)} and \textbf{Au-NP} systems, we applied the HDBSCAN* \cite{McInnes2017} clustering algorithm set up with \texttt{min\_cluster\_size=80} for the former and  \texttt{min\_cluster\_size=150} for the latter, obtaining 7 and 9 environments, respectively.
We used soft-clustering to assign the point classified as noise to their closer cluster.
From the cluster transition probability matrix (see Figure S1a,b), we found the relations within the environments \textit{via} Hierarchical Clustering algorithm.
Then, merging the ones closer than 1 in terms of the chosen metrics (\texttt{correlation}) and linkage (\texttt{average}), we obtained 6 and 7 macro-clusters respectively for \textbf{Cu(211)} and \textbf{Au-NP} systems.
Regarding the \textbf{TIP4P/Ice} system, we followed a slightly different procedure: indeed, as clear from the PCA of the SOAP spectra reported in Figure \ref{fig:04}a, there are no clear density-based patterns, and HDBSCAN* failed to assign meaningful clusters, as shown in Figure S3.
Thus, instead of HDBSCAN* clustering algorithm, we employed a Gaussian Mixture Model (GMM)\cite{reynolds2009} setting the number of clusters to three, without merging clusters \textit{a posteriori} bust still applying HC being interested in cluster relations.
Then, for all the systems, we compute the clusters' fluxes, \textit{i.e.} the number of particles going from one cluster to another, following the cluster assignment along the trajectory.
The fluxes are visualized as chord diagrams in Figures \ref{fig:01}c, \ref{fig:03}a, and \ref{fig:04}a. The width of the arcs represents the total number of transitions experienced by each cluster during the simulation, including both self-transitions and those to other clusters.
The chords linking the clusters depict their interconnections, with the extension of the chord's base indicating the amount of particles exchanging between connected clusters.
The color of the chords indicates the dominant direction of particle transfer between clusters.
Then, normalizing the flux matrices on each row, we obtained the transition probability, reported in Figures \ref{fig:01}c, \ref{fig:03}a and \ref{fig:04}a.

\subsection*{LENS analysis}
We compute the $\delta_i(t)$ signals for all the systems following a similar procedure reported in Crippa \textit{et al.}\cite{crippa2022de}, reducing the noise by using a Savitzky–Golay \cite{savitzky1964} filter (as implemented in the SciPy python package  \cite{virtanen2020}).
Each $\delta_i$(t) signal is smoothed using a common polynomial order parameter of $p=2$ and a time-window of 20 frames in the crystalline \textbf{Cu(211)} surface, 100 frames for both the water/ice interface and the $\textbf{Au-NP}$ system.
After the noise reduction, the clustering of the $\delta_i$ data is performed: in the case of \textbf{Cu(211)}, the clustering thresholds are set as previously \cite{crippa2022de} while for both the \textbf{Au-NP} and the \textbf{TIP4P/Ice} systems are set by means of KMeans algorithm \cite{lloyd1982} implemented in SciPy python package.\cite{virtanen2020}
The KMeans algorithm requires the definition of the number of clusters as an input: in both cases of  gold nanoparticle and ice/water interface, we set four and three clusters respectively, according to the number of macro clusters previously found.\cite{crippa2022de}
Knowing the cluster assignment, we compute the cluster fluxes, \textit{i.e.} the number of particles going from one cluster to another, for each system.
The fluxes are reported as chord diagrams of Figures \ref{fig:01}d, \ref{fig:03}b and \ref{fig:04}b, representing the data as reported above.
Then, normalizing the flux matrices on each row, we obtain the transition probability, reported in Figures \ref{fig:01}d, \ref{fig:03}b, and \ref{fig:04}b.

\subsection*{SOAP\&LENS combined analysis}

The combined SOAP\&LENS descriptor is obtained by following the procedure illustrated in Figure \ref{fig:01}e and explained in detail in the Results section. 
The SOAP power spectrum of each particle $i$ at every time step $t$ ($\textbf{p}^t_i$) is combined with the subsequent LENS scalar value ($\delta^{t+\Delta t}_i$), obtaining a new vector $\boldsymbol{\chi}_i^{t}=(\textbf{p}_i^{t}, \delta_i^{t+\Delta t})$.
Each SOAP power spectra are normalized on their norm, while the LENS scalar is intrinsically normalized within zero (no neighborhood changes) and one (the whole neighborhood changes).
In this way, while retaining different information and having two distinct mathematical forms (a high dimensional vector and a scalar), the two components have the same "weight" in the dataset.
This procedure, when iterated throughout the entire trajectory, results in a new dataset including $N = N_{particle} \times N_{frames}$ vectors.
Each vector contains $n+1$ components: $n$ components representing the SOAP power spectrum and 1 component representing the LENS value.
Starting from this $\boldsymbol{\chi}_i^{t}$ representation of the particle local environments, we follow the same \textit{bottom-up} procedure, described above, applied to the pure SOAP dataset.
To highlight the real effect of the LENS component, avoiding biased results, we performed the \textit{bottom-up} analysis by using the same parameters.
Indeed, upon applying PCA to the SOAP\&LENS dataset of each system, we considered the first n-PCA components to match the PCA variance retained in the SOAP analysis, as reported in Table S1.
We apply the clustering algorithm (both HDBSCAN* and GMM) to this new reduced dataset, by using the same parameters (\texttt{min\_cluster\_size=80} for the \textbf{Cu(211)} and  \texttt{min\_cluster\_size=150} for the \textbf{Au-NP} and \texttt{n\_component}=3 for the \textbf{TIP4P/Ice}), and then the HC dendrogram cutting under the same conditions \textit{i.e.} closer than 1 in terms of the chosen metrics (\texttt{correlation}) and linkage (\texttt{average}).

\section*{Data availability}
Details on the molecular models and on the MD simulations, and additional simulation data are provided in the Supporting Information.
Complete data for the simulation and analysis performed in this work are available at: \url{https://github.com/GMPavanLab/StrDynRel} (this temporary folder will be replaced with a definitive Zenodo archive upon acceptance of the final version of this paper).

\section*{Acknowledgements}
G.M.P. acknowledges the support received by the European Research Council (ERC) under the European Union’s Horizon 2020 research and innovation program (Grant Agreement no. 818776 - DYNAPOL) and by the Swiss National Science Foundation (SNSF Grant IZLIZ2$\_183336$).

\section*{Competing interests statement} 
The authors declare no competing interests.

\bibliography{bibliography}
\end{document}


\maketitle
\begin{table}[]
\begin{tabular}{|l|l|l|l|l|l|}
\hline
SYS                & r$_{cut}$ $\si{\angstrom}$ &\shortstack{ n-PC \\ SOAP }& \shortstack{ PC var \\ SOAP }      & \shortstack{ n-PC \\ SOAP\&LENS } & \shortstack{ PC var \\ SOAP\&LENS }   \\ \hline
\textbf{Cu(211)}   & 6             & 4         & \textgreater 99.5\% & 5               & \textgreater 99.5\% \\ \hline
\textbf{Au-NP}     & 4.48          & 3         & \textgreater 99.5\% & 4               & \textgreater 99.5\% \\ \hline
\textbf{TIP4P/Ice} & 6             & 8         & \textgreater 99\%   & 3               & \textgreater 99\%   \\ \hline
\end{tabular}
\caption{Details about SOAP and SOAP\&LENS analysis}
\end{table}

\begin{figure}
    \centering
    \includegraphics[width=\columnwidth]{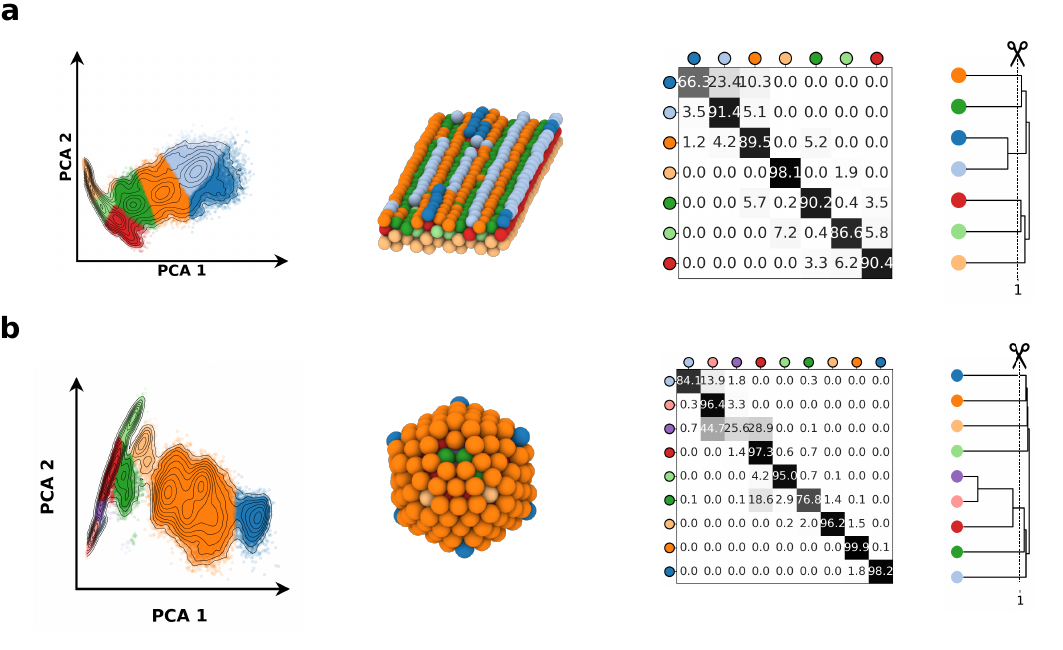}
    \caption{SOAP analysis. Left: First to PC of the PCA on SOAP dataset and a representative snapshot, of \textbf{Cu(211)} (a) and \textbf{Au-NP} (b), colored accordingly to HDBSCAN* clustering (\texttt{min\_cluster\_size=80} and \texttt{min\_cluster\_size=150}, respectively). Right: Transition probability matrix obtained from the particles transition within the trajectory: each out-of-diagonal cell represents the probability of a particle in a certain cluster (row) to transit to another cluster (column). HC-based dendrogram of the transition matrix rows: the rows are clustered based on their distance (\texttt{correlation} metrics) and grouped considering the average of the new clusters (\texttt{average} linkage). The dendrogram is cut at correlation equal to 1.}
    \label{fig:s1}
\end{figure}

\begin{figure}
    \centering
    \includegraphics[width=\columnwidth]{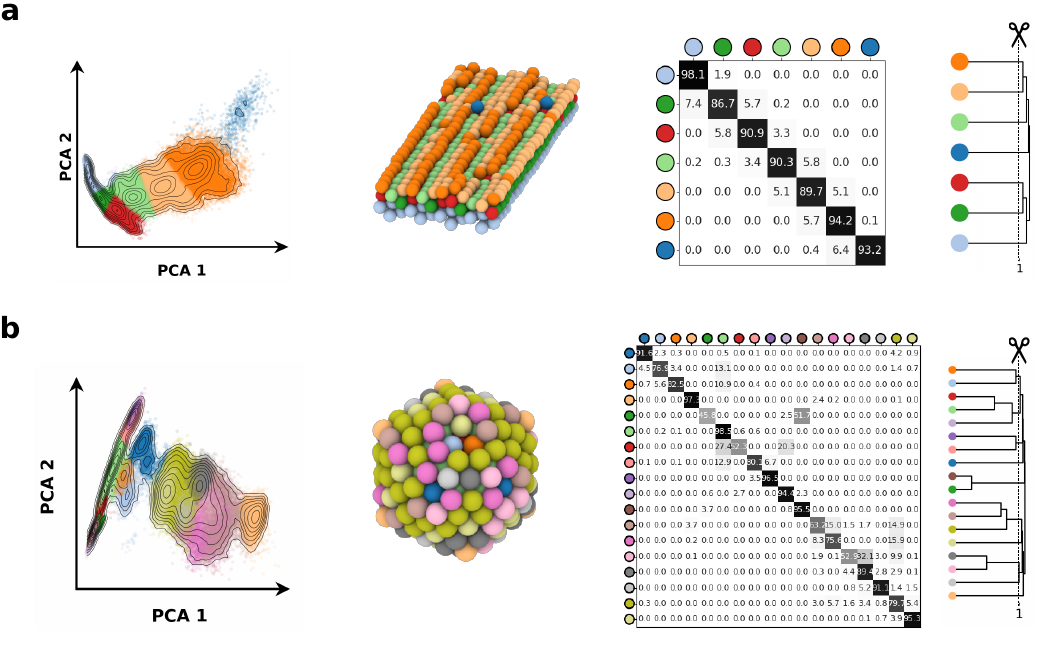}
    \caption{SOAP\&LENS analysis. Left: First to PC of the PCA on SOAP\&LENS dataset and a representative snapshot, of \textbf{Cu(211)} (a) and \textbf{Au-NP} (b), colored accordingly to HDBSCAN* clustering (\texttt{min\_cluster\_size=80} and \texttt{min\_cluster\_size=150}, respectively). Right: Transition probability matrix obtained from the particles transition within the trajectory: each out-of-diagonal cell represents the probability of a particle in a certain cluster (row) to transit to another cluster (column). HC-based dendrogram of the transition matrix rows: the rows are clustered based on their distance (\texttt{correlation} metrics) and grouped considering the average of the new clusters (\texttt{average} linkage). The dendrogram is cut at correlation equal to 1.}
    \label{fig:s2}
\end{figure}

\begin{figure}
    \centering
    \includegraphics[width=\columnwidth]{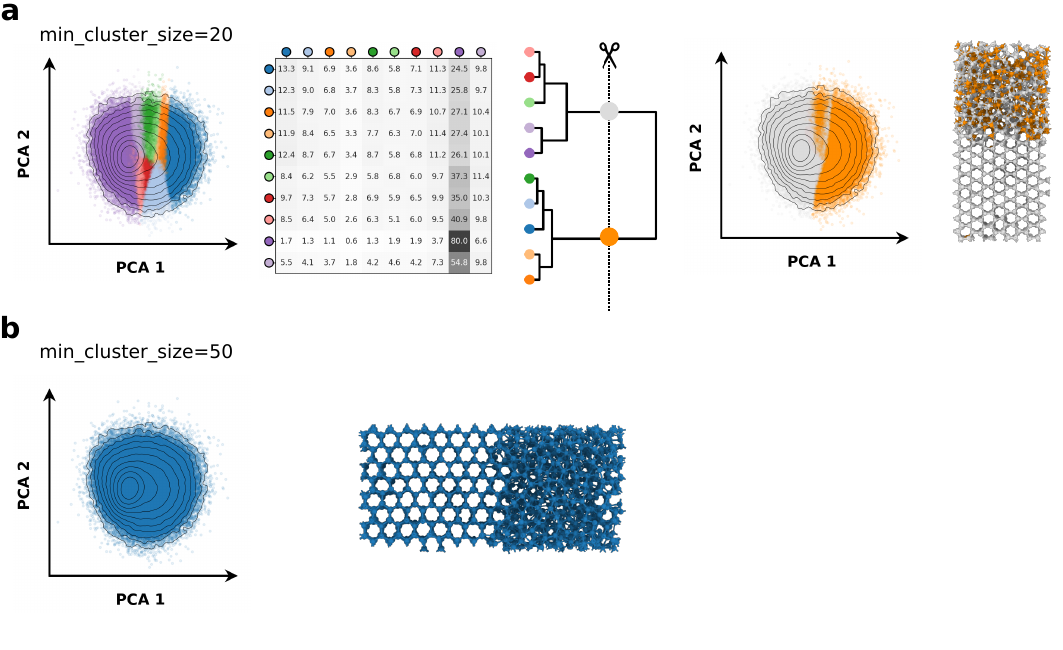}
    \caption{SOAP additional analysis for the liquid/ice interface. Left: First to PC of the PCA on SOAP dataset of \textbf{TIP4P/Ice} system, colored accordingly to HDBSCAN* clustering applied two different parameters: \texttt{min\_cluster\_size=20} (a) and \texttt{min\_cluster\_size=50} (b). Right: for (a), transition probability matrix obtained from the particles transition within the trajectory: each out-of-diagonal cell represents the probability of a particle in a certain cluster (row) to transit to another cluster (column). HC-based dendrogram of the transition matrix rows: the rows are clustered based on their distance (\texttt{correlation} metrics) and grouped considering the average of the new clusters (\texttt{average} linkage). The dendrogram is cut to obtain two environments, not well-defined as depicted in the snapshot on the right.}
    \label{fig:s3}
\end{figure}

\begin{figure}
    \centering
    \includegraphics[width=\columnwidth]{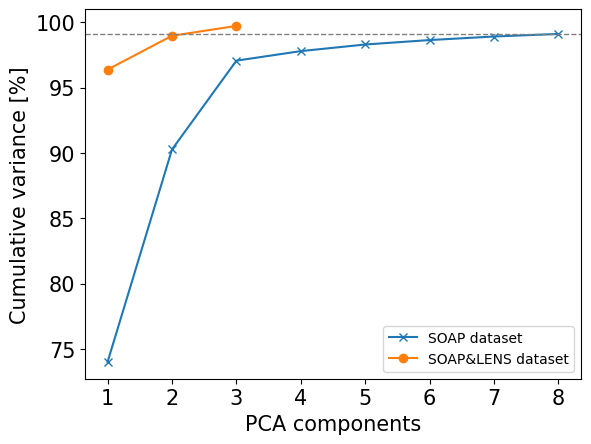}
    \caption{PCA cumulative variance in function of the number of the PCA components, comparing the pure SOAP dataset with the SOAP\&LENS combined one. In dotted gray a line corresponding to the variance reached by the 8th PCA components of the SOAP dataset.}
    \label{fig:s4}
\end{figure}